\documentclass[aip,amsmath,amssymb,reprint,]{revtex4-2}

\usepackage{graphicx}
\usepackage{dcolumn}
\usepackage{bm}

\usepackage[utf8]{inputenc}
\usepackage[T1]{fontenc}
\usepackage{mathptmx}
\usepackage{etoolbox}
\usepackage{xcolor}

\makeatletter
\def\@email#1#2{%
 \endgroup
 \patchcmd{\titleblock@produce}
  {\frontmatter@RRAPformat}
  {\frontmatter@RRAPformat{\produce@RRAP{*#1\href{mailto:#2}{#2}}}\frontmatter@RRAPformat}
  {}{}
}%
\makeatother

\begin{document}


\preprint{}

\title{Geometry-induced self-excited dynamo in a regular tetrahedron}

\author{Akira Kageyama}
\affiliation{Graduate School of System Informatics, Kobe University, Kobe 657-8501, Japan}
\email{kage@port.kobe-u.ac.jp}

\date{\today}

\begin{abstract}
We present a rotation-free magnetohydrodynamic dynamo driven by laminar thermal convection in a regular tetrahedral cavity. The tetrahedral boundaries organize the convective flow into a robust pattern of helical convection cells without global rotation or turbulence. Direct numerical simulations demonstrate exponential amplification of a weak seed magnetic field followed by nonlinear saturation. In the reference calculation, the saturated magnetic energy exceeds the kinetic energy. The velocity field develops $D_4$ dihedral symmetry, while the self-generated magnetic field exhibits a corresponding signed $D_4$ symmetry, including antisymmetry under $\pi$ rotations about the two horizontal axes. Analysis of the velocity and magnetic-field structures reveals a closed induction cycle sustained by geometry-induced helical convection. This system provides a conceptually simple setting for isolating and understanding the fundamental physical processes underlying magnetohydrodynamic dynamo action.
\end{abstract}

\maketitle

\section{Introduction}

Magnetic-field generation by electrically conducting flows is a fundamental problem in magnetohydrodynamics (MHD) and plasma physics. In the Sun and other stars, dynamo action driven by plasma flows maintains large-scale magnetic fields and produces diverse forms of magnetic activity~\cite{Jones2010-vi,Charbonneau2020-re,Brun2017-ba}. The geomagnetic field is likewise maintained by a self-induction process, although the electrically conducting medium in the Earth's outer core is liquid metal rather than plasma. Despite this difference, both systems are governed by the basic MHD process through which fluid motion generates and sustains magnetic fields. A central challenge in geodynamo theory~\cite{Kono2007-xe,Roberts2013-nd} is to explain the origin, morphology, and variability of the geomagnetic field. The observed geomagnetic properties emerge from a highly complex MHD convection system driven by thermal and compositional buoyancy and shaped by rotation, turbulence, boundary layers, core--mantle boundary heterogeneity, and the presence of the inner core~\cite{Dormy2007-aw,Nimmo2015-ni,Jones2015-xr,Sumita2015-so,Dormy2025-yx}. This complexity motivates simplified models that isolate elementary dynamo mechanisms applicable across different conducting-fluid systems.

A representative simplified approach is provided by laboratory dynamo experiments using liquid metals realized in diverse settings, including externally forced flows and duct flows~\cite{Petrelis2025-le,Gailitis2000-ev,Muller2008-dt,Ravelet2012-mx}.
Precession-driven dynamo experiments are also being pursued~\cite{Stefani2025-my}.
Conceptually related systems based on conducting solids or disks rather than fluid flows have likewise been explored~\cite{Alboussiere2022-fl,Alejandro_Avalos-Zuniga2023-bi}.
Despite their different realizations, these systems all require mechanisms that systematically stretch and reorient magnetic field lines.
In fluid dynamos, genuinely three-dimensional helical motion provides one important route to such induction~\cite{Moffatt2016-qk}.
Kinetic helicity is not, however, a necessary condition for dynamo action; steady three-dimensional flows with pointwise vanishing helicity can also generate magnetic fields~\cite{Rasskazov2018-qn,Andrievsky2019-hm}.

Here we take a numerical-simulation approach and investigate whether a simple rotation-free fluid system can sustain dynamo action.
We show that laminar thermal convection in a regular tetrahedral cavity spontaneously organizes into a robust three-dimensional helical flow and sustains a self-excited dynamo. In the reference calculation, the saturated magnetic energy exceeds the kinetic energy of the convection.
Our aim is not to establish helicity as a necessary condition for dynamo action, but to determine how tetrahedral geometry organizes thermal convection and how the resulting flow closes a self-sustaining induction cycle.
Unlike the geophysically motivated rotating spherical-shell systems considered in our previous simulations~\cite{Kageyama1997-yt,Kageyama1999-ca,Kageyama2008-gy}, the present model is not intended as a realistic representation of a natural MHD dynamo.
Rather, it is a simple nonlinear dynamical system designed to isolate MHD dynamo mechanisms in the absence of global rotation and its associated effects, such as the Coriolis force and Ekman boundary layers~\cite{Dormy2007-aw}.

We do not claim that the regular tetrahedron is unique or optimal for dynamo action.
Among the Platonic solids, it has the fewest faces and the smallest interior dihedral angle, which led us to expect particularly clear boundary-induced organization of the flow.
Its high symmetry facilitates analysis of the velocity and magnetic-field structures, and its vertices, edges, and faces can be represented without geometric approximation on a uniform Cartesian grid; see Appendix~\ref{260402113313}.
Other polyhedral or distorted geometries may likewise organize convection into dynamo-capable flows, but such extensions remain to be investigated.

Thermal convection has been studied in a variety of nonstandard container geometries~\cite{Das2017-yl,Abdulkadhim2021-kr,Dittko2013-ak}.
Fontana et al.~\cite{Fontana2024-oo} recently examined convection in Platonic solids, including the regular tetrahedron, although their tetrahedral configuration differs from the one considered here.
Large-scale circulations are also known to emerge in vigorous thermal convection at high Rayleigh numbers, even in standard geometries such as rectangular boxes, cylinders, and spherical shells~\cite{Krishnamurti1981-hh,Qiu2001-kf,Chen2024-yy}.
In contrast, the present study focuses on a steady laminar helical flow capable of sustaining dynamo action.

\section{Model}
\begin{figure}[ht]   \centering
  \includegraphics[width=0.6\columnwidth]{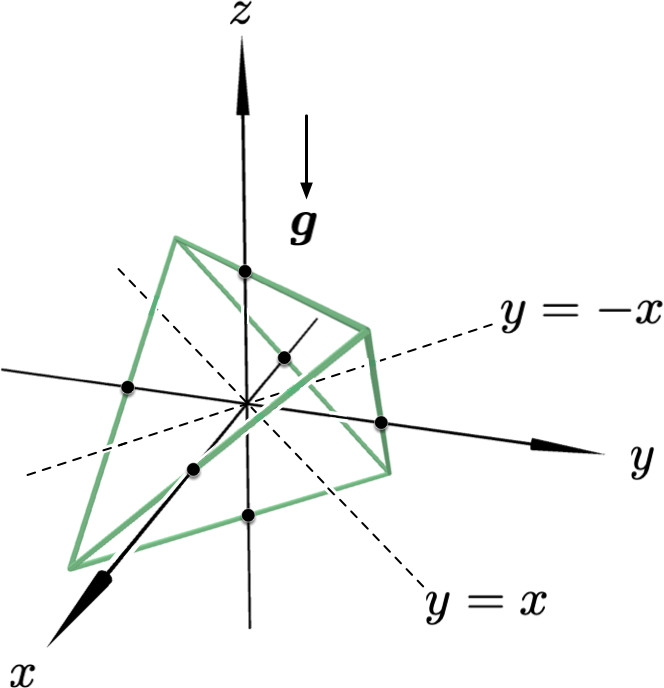}
  \caption{Regular tetrahedral container and coordinate system.
Gravity points in the \(-z\) direction, and the fluid is heated from below.
The tetrahedron is oriented so that one pair of opposite edges is horizontal.
Each coordinate axis connects the midpoints of one pair of opposite edges.}
  \label{260414110438}
\end{figure}
We place a regular tetrahedron in a uniform gravitational field and orient it so that one pair of opposite edges is horizontal; see Fig.~\ref{260414110438}.
The tetrahedron has height \(1\,\mathrm{m}\) between the horizontal edges and edge length \(\sqrt{2}\,\mathrm{m}\).
The $z$-axis is aligned with the line connecting the midpoints of the pair of opposite horizontal edges, and the $x$- and $y$-axes are taken along the lines connecting the midpoints of the other two pairs of opposite edges.

We solve the following form of the compressible MHD equations.
\begin{align} 
  \frac{\partial\rho}{\partial t} &=-\nabla\cdot\bm{f},     \label{240305173145a} \\
  \frac{\partial\bm{f}}{\partial t} &=-\nabla\cdot(\bm{v}\bm{f})-\nabla p + \bm{j}\times \bm{b} + \rho\bm{g}+\mu\left\{\nabla^2\bm{v}+(1/3)\nabla(\nabla\cdot\bm{v})\right\},  \label{240305173145b} \\
  \frac{\partial p}{\partial t}   &= -\bm{v}\cdot\nabla p - \gamma p\nabla\cdot\bm{v}+(\gamma-1)\Phi, \label{240305173145c} \\
  \frac{\partial\bm{b}}{\partial t}   &= \nabla\times\left(\bm{v}\times\bm{b}\right) + \frac{\eta}{\mu_0}\nabla^2\bm{b}, \label{240305173145d} 
\end{align} 
Here $\bm{j}=\nabla\times\bm{b}/\mu_0$ is the electric current density.
$\rho$ (mass density), $\bm{f}=\rho\bm{v}$ (mass flux), $p$ (pressure), and $\bm{b}$ (magnetic field) are fundamental field variables and $\bm{v}$ (velocity) and $T$ (temperature) are subsidiary fields.
We assume the ideal gas equation of state with the ratio of specific heats $\gamma = 5/3$.
$\Phi=\Phi_\mu +\Phi_\kappa + \Phi_\eta$ is the dissipation function with dissipation parameters $\mu$ (viscosity), $\kappa$ (thermal conductivity), and $\eta$ (resistivity);
$\mu$, $\kappa$, and $\eta$ are constants.
$\Phi_\mu$, $\Phi_\kappa$, and $\Phi_\eta$ denote viscous heating, thermal conduction, and Ohmic heating, respectively.
$\bm{g} = -g\hat{\bm{e}}_z$ is the gravitational acceleration,
where $\hat{\bm{e}}_z=(0,0,1)$ is the unit vector in the $z$ direction.
We solve the equations in dimensional form, using SI units for all physical quantities.

Prandtl number $Pr =  \nu/\chi$, 
and magnetic Prandtl number $Pm = \mu_0\nu/\eta$
are defined through the thermal diffusivity $\chi =\kappa/(c_p \rho)$
and the kinematic viscosity $\nu = \mu/\rho$, where $c_p$ is the specific heat at constant pressure.

The initial condition is given by a hydrostatic polytropic state with a parameter $\alpha$ and the temperature gradient constant $\beta$ as
\begin{align}
   p(z) &= p(x,y,z,t=0) = p_0 (1-\beta z)^{\alpha+1},  \label{240314175003a} \\
   \rho(z) &= \rho(x,y,z,t=0) = \rho_0 (1-\beta z)^\alpha,   \label{240314175003b} \\
   T(z) &= T(x,y,z,t=0) = T_0 (1-\beta z),   \label{240314175003c}
\end{align}
where $p_0$, $\rho_0$, and $T_0$ are values at $z=0$.
Two constants $\alpha$ and $\beta$ are related as $ (\alpha+1) \beta p_0 = \rho_0 g$.
The initial velocity and magnetic fields are $\bm{v}=\bm{b}=0$.
Random weak perturbations on the pressure and weak seed magnetic field are added.

In compressible convection, the local Rayleigh number generally varies with depth owing to the density stratification and the resulting depth dependence of the diffusivities~\cite{Spiegel1965-qc}:
\begin{equation} \label{250930171302} 
   Ra = \frac{g d^4}{\nu\chi}\frac{T_0}{T}
   		\left(
			\beta - \beta_\mathrm{a}
		\right),
\end{equation}
where $d=1\,\mathrm{m}$ is the height of the regular tetrahedron, the distance between the two horizontal edges,
and $\beta_\mathrm{a}=g/(c_\mathrm{p} T_0)$ is the adiabatic temperature gradient.
In the present study, we set $\alpha=1/2$.
Since $Ra \propto T(z)^{2\alpha-1}$, this choice makes $Ra$ independent of $z$.
Thus, $\alpha=1/2$ is a convenient choice for keeping $Ra$ uniform in $z$ and is not regarded as essential to the dynamo mechanism.
We use $\rho_0=1.0\times 10^3\,\mathrm{kg\,m^{-3}}$,
$T_0=3\times 10^2\,\mathrm{K}$,
$Pr=1.0$, and $Pm=5.0$.
The value of $Pm$ is evaluated at $z=0$.
The choice $Pm=5$ provides a numerically accessible dynamo regime and is not intended to represent the much smaller magnetic Prandtl number of the Earth's core.
The dimensional transport coefficients used in the simulations are the dynamic viscosity $\mu=3.471\times10^2\,\mathrm{Pa\,s}$, the thermal conductivity $\kappa=1.432\times10^7\,\mathrm{W\,m^{-1}\,K^{-1}}$, and the electrical resistivity $\eta=8.724\times10^{-8}\,\Omega\,\mathrm{m}$.
The gravitational acceleration is set to $g=9.8\times10^5\,\mathrm{m\,s^{-2}}$ to obtain the prescribed Rayleigh number for the chosen dimensional scales.
These dimensional parameter values provide a convenient numerical realization of the chosen dimensionless parameters and are not intended to represent a realistic laboratory setup.

Numerically, we find the critical Rayleigh number to be $Ra_\mathrm{c}=43000$.
For comparison, a conventionally oriented cube with horizontal top and bottom faces and adiabatic sidewalls~\cite{Mizushima2003-gs} has \(Ra_\mathrm{c}=3389\).
The flow is weakly stratified and has a low Mach number.
The density stratification is $\rho_\text{top}/\rho_\text{bottom}=0.936$,
and the maximum Mach number of the resulting convection is $M = 1.20\times10^{-2}$.
Although temperature-dependent density variations are essential for thermal buoyancy, we expect the dynamo to persist in the Boussinesq limit, where buoyancy is retained but other compressible dynamics are neglected; this expectation has not been tested directly, and the fully compressible equations are used here for consistency with our numerical framework.

The four triangular surfaces of the tetrahedron are impermeable and no-slip for the fluid.
Fixed-temperature boundary conditions are imposed, with $T=T_0(1-\beta z)$.

The tetrahedron is circumscribed by a cube with edge length $1\,\mathrm{m}$.
The fluid equations are solved only inside the tetrahedron, whereas the induction equation is solved in a slightly larger cubic domain with edge length $1+2\Delta$, where $\Delta$ is the grid spacing described below.
The exterior region is treated as a motionless conductor with the same magnetic diffusivity as the fluid, and periodic boundary conditions for the magnetic field are imposed on the outer cubic boundary.

We use a fourth-order explicit Runge--Kutta method for time integration and a second-order central finite-difference method for spatial discretization on a collocated grid.
The simulations are performed on a uniform Cartesian grid with $\Delta x = \Delta y = \Delta z = \Delta$.
According to the arrangement of the coordinate axes and the regular tetrahedron shown in Fig.~\ref{260414110438},
the tetrahedral vertices, edges, and faces are placed exactly on grid points.
This integer-lattice embedding allows a second-order central finite-difference scheme and explicit fourth-order Runge--Kutta time-stepping while imposing the boundary conditions directly on grid points; details of the grid-conforming placement and boundary implementation are given in Appendix~\ref{260402113313}.
The reference calculation uses a grid of $N_x\times N_y\times N_z=80^3$.
Note that with the second-order central-difference operators on the uniform Cartesian grid, the discrete identities
\(\nabla_h\cdot(\nabla_h\times\bm{a})=0\) for any discrete vector field \(\bm{a}\) and
\(\nabla_h\cdot\nabla_h^2\bm{b}=\nabla_h^2(\nabla_h\cdot\bm{b})\) both hold.
Thus, an initially solenoidal magnetic field remains solenoidal up to round-off error.

\section{Results}
\begin{figure}[t]  \centering
  \includegraphics[width=\columnwidth]{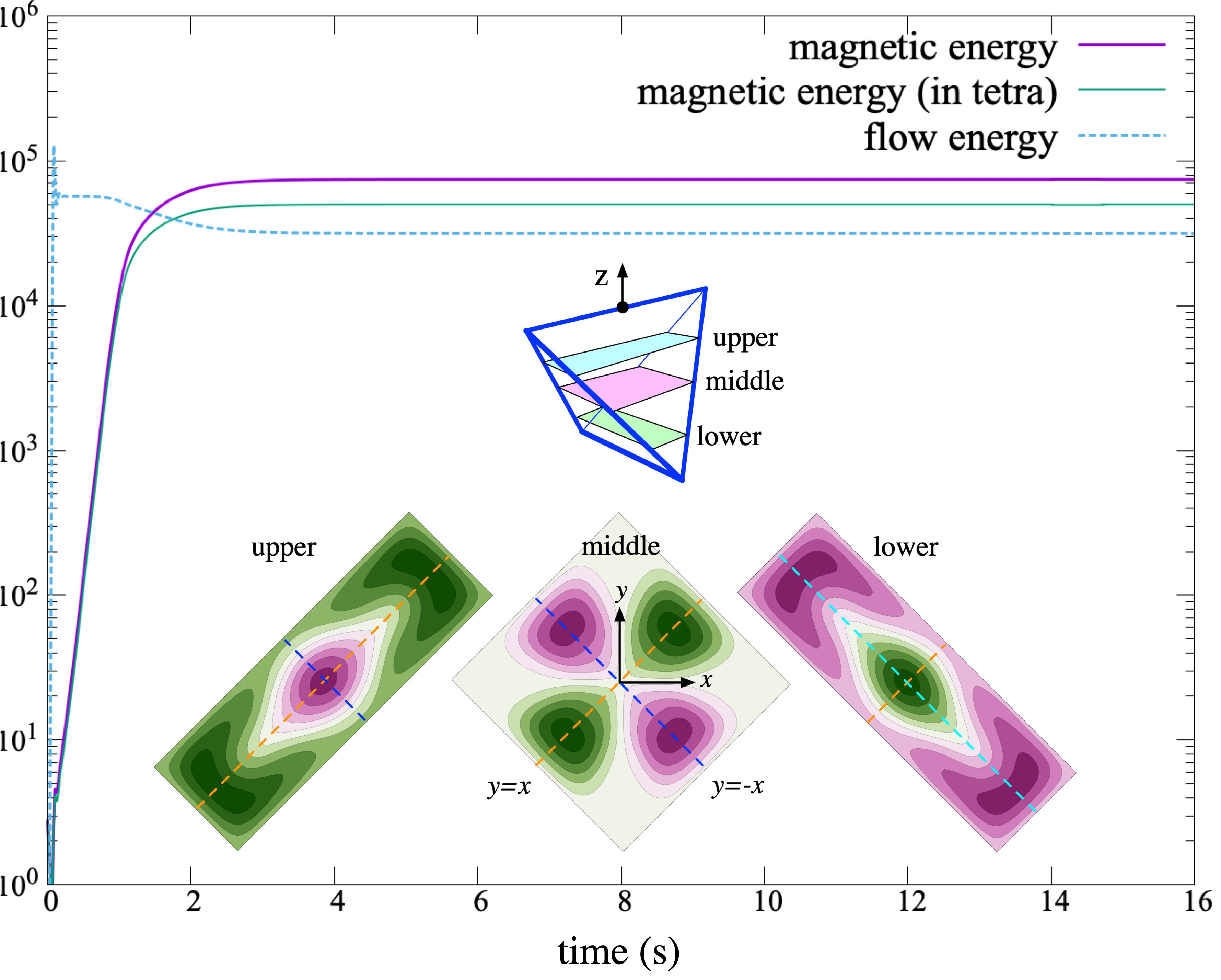}
  \caption{Time evolution of kinetic and magnetic energies for \(Ra=10\,Ra_c\).
The seed magnetic field grows exponentially and reaches a saturated state.
In this reference calculation, the saturated magnetic energies integrated over the entire computational domain and over the tetrahedron both exceed the kinetic energy.
The final state is steady and laminar.
The lower panels show horizontal cross sections of the vertical velocity component \(v_z\) in the steady state, at the three heights indicated in the inset.
Positive and negative values are shown in shades of purple and green, respectively.}
  \label{260420095816}
\end{figure}
We focus on a representative case at $Ra=10\,Ra_c$ with $Pr=1$ and $Pm=5$.
At this parameter value, the convection settles into a steady laminar state.
Fig.~\ref{260420095816} shows the time development of the convection kinetic energy and the magnetic energies.
Since the magnetic field penetrates outside the regular tetrahedron, both the total magnetic energy over the entire domain and the magnetic energy confined within the tetrahedron are plotted.

The energies plotted in Fig.~\ref{260420095816} are defined by
\begin{align}
E_K & = \int_{\mathcal{T}} \frac{1}{2}\rho |\bm{v}|^2\,dV, \\
E_M^{(\mathcal{T})} & = \int_{\mathcal{T}} \frac{|\bm{b}|^2}{2\mu_0}\,dV, \\
E_M^{(\mathcal{D})} & = \int_{\mathcal{D}} \frac{|\bm{b}|^2}{2\mu_0}\,dV,
\end{align}
where $\mathcal{T}$ denotes the tetrahedral fluid domain and $\mathcal{D}$ denotes the entire computational domain.

To estimate the dynamo threshold for this convective flow, we performed additional calculations on the $80^3$ grid at $Pm=1.9$, $2.0$, and $2.1$, with the other parameters fixed at their reference values.
Defining the kinematic growth rate $\gamma$ by $E_M^{1/2}\propto\exp(\gamma t)$, we obtain $\gamma=-0.722\,\mathrm{s^{-1}}$ at $Pm=1.9$ and $\gamma=0.089\,\mathrm{s^{-1}}$ at $Pm=2.1$, while the magnetic field at $Pm=2.0$ decays only slowly.
Linear interpolation between the two bracketing growth rates gives $Pm_c\simeq2.08$, indicating that the reference case at $Pm=5$ lies well above the estimated dynamo onset.
A systematic investigation of the dependence on $Ra$ and $Pr$ is left for future work.

After a short overshoot, the kinetic energy settles onto its initial plateau by $t \simeq 0.3\,\mathrm{s}$.
Using the mass-weighted rms velocity $U_\mathrm{rms}=(2E_K/\int_{\mathcal{T}}\rho\,dV)^{1/2}$ and the values $E_K=5.71\times10^4\,\mathrm{J}$, $V_{\mathcal{T}}=1/3\,\mathrm{m^3}$, and $\rho\simeq1.0\times10^3\,\mathrm{kg\,m^{-3}}$ at $t=0.3\,\mathrm{s}$, we obtain $U_\mathrm{rms}\simeq18.5\,\mathrm{m\,s^{-1}}$.
With $d=1\,\mathrm{m}$, the corresponding Reynolds and magnetic Reynolds numbers are $Re=U_\mathrm{rms}d/\nu=53.3$ and $Rm=U_\mathrm{rms}d/(\eta/\mu_0)=Pm\,Re\simeq267$, respectively.
During this early stage, the magnetic field grows exponentially, 
but its energy remains too small to significantly affect the flow. 
For the reference case at $Pm=5$, a least-squares fit of $(1/2)\ln E_M$ as a function of time over $0.50\leq t\leq0.60\,\mathrm{s}$ gives $\gamma=5.01\,\mathrm{s^{-1}}$.
In this reference calculation, both magnetic energies subsequently exceed the kinetic energy, while the kinetic energy decreases from its initial saturation level in the absence of magnetic fields, resulting in crossings of the curves.
The system eventually reaches a steady nonlinear saturated state. 

The decrease in kinetic energy after the magnetic field becomes dynamically significant indicates Lorentz-force feedback on the convection.
In the steady saturated state, magnetic-energy production by the flow balances Ohmic dissipation, while the buoyancy power input is ultimately dissipated through viscous and Ohmic losses, with the Lorentz force mediating energy transfer between the kinetic and magnetic reservoirs~\cite{Buffett2002-oy}.
Magnetic-to-kinetic energy ratios substantially greater than unity are not unusual in numerical geodynamo models~\cite{Kono2002-mb}.
The distinctive feature of the present result is therefore not super-equipartition itself, but its occurrence in a rotation-free, steady laminar dynamo.
In the absence of the Coriolis force, the steady momentum balance involves pressure-gradient, buoyancy, inertial, viscous, and Lorentz forces.
Because the Lorentz force depends on the spatial structure of the magnetic field as well as its energy, equality of the integrated magnetic and kinetic energies is not required.
The saturated energy ratio is therefore not expected to be universal and, as shown by the exterior-domain test reported below, depends on the configuration.

We repeated the simulations with several independent random initial perturbations. 
In all cases, the flow and magnetic fields converged to the same saturated state with the symmetry described below.
The global polarity of the magnetic field, corresponding to \(\bm{b}\to-\bm{b}\), emerged spontaneously from the random initial perturbations.
The magnetic field persists well beyond one magnetic diffusion time, $\tau_\eta = \mu_0 d^2/\eta \simeq 14\,\mathrm{s}$, indicating nonlinear dynamo saturation rather than a transient amplification.

To assess spatial convergence, we repeated the reference calculation on a $120^3$ grid.
For the $80^3$ and $120^3$ grids, the kinematic growth rates are $5.01$ and $4.96\,\mathrm{s^{-1}}$, respectively.
The corresponding saturated magnetic energies are $E_M^{(\mathcal{D})}=7.46\times10^4$ and $7.47\times10^4\,\mathrm{J}$ over the entire computational domain, and $E_M^{(\mathcal{T})}=5.00\times10^4$ and $4.98\times10^4\,\mathrm{J}$ within the tetrahedron.
The magnetic-field structure and symmetry are essentially unchanged between the $80^3$ and $120^3$ calculations; over the entire computational domain, the normalized energy fraction of the component with the signed $D_4$ symmetry discussed in Sec.~\ref{sec:symmetries} is $0.999$ for the $120^3$ magnetic field.
The relative differences in the growth rate, total-domain magnetic energy, and tetrahedral magnetic energy are $0.96\%$, $0.18\%$, and $0.51\%$, respectively, indicating convergence of these quantities with spatial resolution.

To examine the influence of the exterior magnetic boundary, we enlarged the edge length of the periodic outer cube by $35\%$ while keeping the tetrahedron and grid spacing fixed.
The kinematic growth rate decreased from $5.01$ to $4.48\,\mathrm{s^{-1}}$.
The saturated values of $(E_M^{(\mathcal{D})},E_M^{(\mathcal{T})},E_K)$ changed from $(7.46,5.00,3.16)\times10^4\,\mathrm{J}$ to $(4.44,3.04,4.32)\times10^4\,\mathrm{J}$.
Thus, the exterior-domain size affects the growth rate and saturation levels, but exponential magnetic-field growth and nonlinear saturation persist when the outer boundary is moved farther from the tetrahedron.
The enlarged exterior domain also yields a less symmetric magnetic-field structure, indicating that the detailed dynamo mode depends on the exterior-domain size.
A detailed analysis of this structural change is left for future work.

In the lower part of Fig.~\ref{260420095816},
we show horizontal cross sections of the $z$ component of the laminar velocity field in the saturated state, taken at the midplane $z=0$ and at $z=\pm 0.25$, above and below it.
We note that in the present orientation, a cross section of the regular tetrahedron by a plane of constant $z$ is a rectangle.
A highly symmetric structure of the flow is evident.

\begin{figure}[ht]   \centering   
  \includegraphics[%
     height=1.0\textheight,%
       width=1.0\hsize,keepaspectratio]%
         {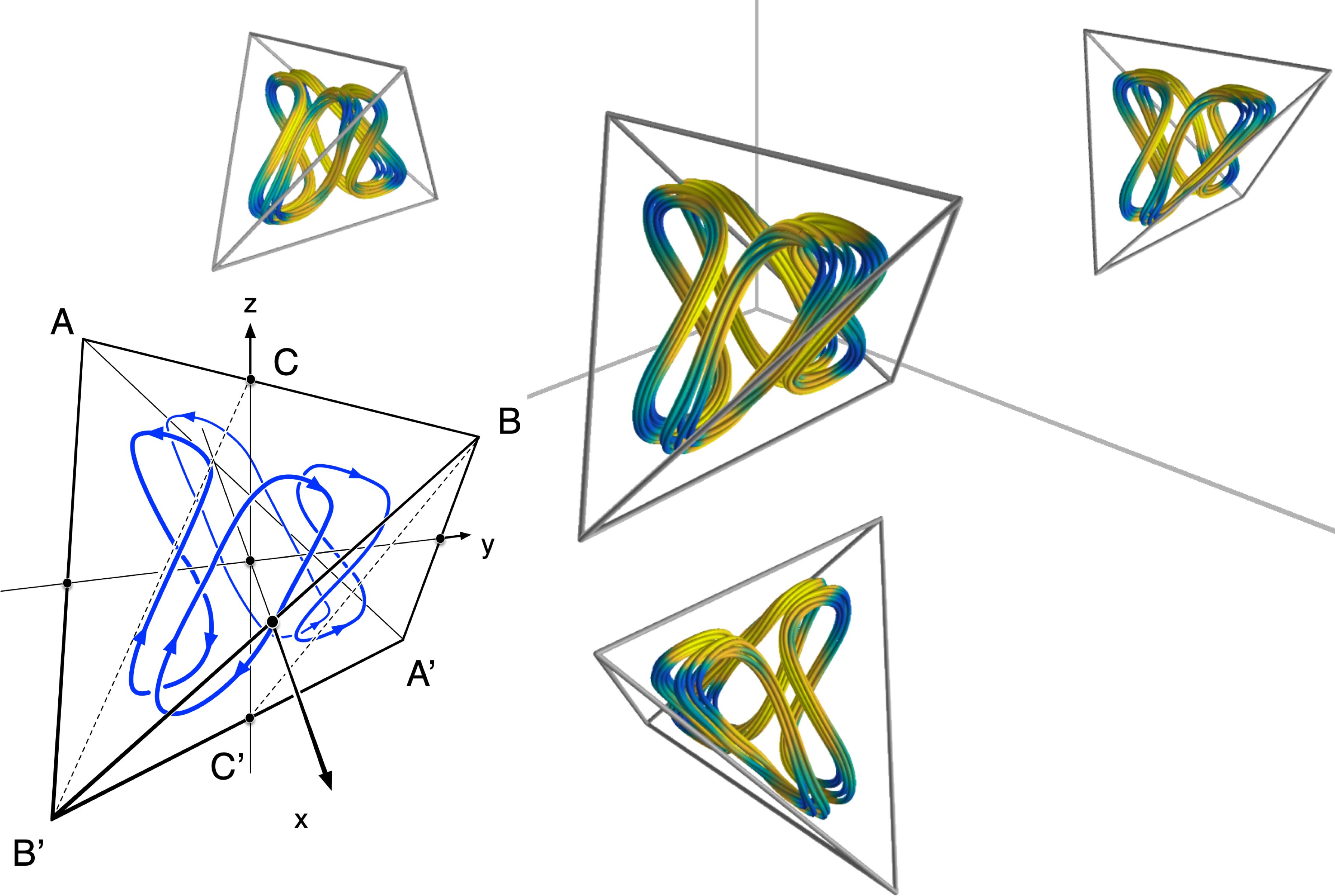}
      \caption{Streamlines of the saturated convective velocity organized into four cells.
The streamlines are seeded at local maxima of the absolute flow helicity in the \(z>0\) part of each convection cell.
The three virtual mirrors help reveal the three-dimensional structure of the streamlines.}
      \label{260319113840}
\end{figure} 
Fig.~\ref{260319113840} shows streamlines of the flow, which reveal the three-dimensional structure of the velocity field.
To aid visual perception, three mutually orthogonal virtual mirrors are introduced on the two side walls and on the bottom plane, revealing the underlying spatial symmetries.
The flow is organized into four convection cells separated by two vertical surfaces at $y=\pm x$. 
The four streamlines, one in each convection cell, exhibit a twisting structure.
One convection cell is the triangular pyramid $CBC'B'$, where the edge $AB$ lies in the plane $y=x$.
The vertices $A'$ and $B'$ are obtained from $A$ and $B$, respectively, by a $\pi$-rotation about the $x$ axis.

We focus on the streamline in the convection cell \(CBC'B'\).
A key point is that this triangular pyramid is invariant under the \(\pi\)-rotation about the \(x\) axis. 
A fluid particle that starts near the bottom vertex $B'$ rises by buoyancy along the median $B'C$. 
After entering the region \(z>0\), it turns and flows horizontally toward the vertex $B$. 
It then turns again and begins to descend toward \(z<0\) along the median $BC'$. 
When this motion is viewed from outside the tetrahedron along the outward normal to the face $CBB'$, the trajectory forms a right-handed helix.

The corresponding lower-half motion descends from \(B\) toward \(C'\), turns toward \(B'\), and then rises again from \(B'\).
This motion is obtained from the right-handed helical flow described above by the \(\pi\)-rotation about the \(x\) axis.
The helical trajectory is therefore also right-handed.
The volume-integrated helicity over the cell $CBC'B'$ is positive.
By contrast, the neighboring cell $CAC'B'$ is related to $CBC'B'$ by reflection symmetry and therefore has negative volume-integrated helicity.
The same symmetry argument shows that the cell $CBC'A'$ has negative helicity, whereas $CAC'A'$ has positive helicity.
Consequently, the volume-integrated kinetic helicity over the entire tetrahedron vanishes, even though the helicity integrated over each individual convection cell is nonzero.

\begin{figure}[ht]   \centering   
  \includegraphics[%
     height=1.0\textheight,%
       width=1.0\hsize,keepaspectratio]%
         {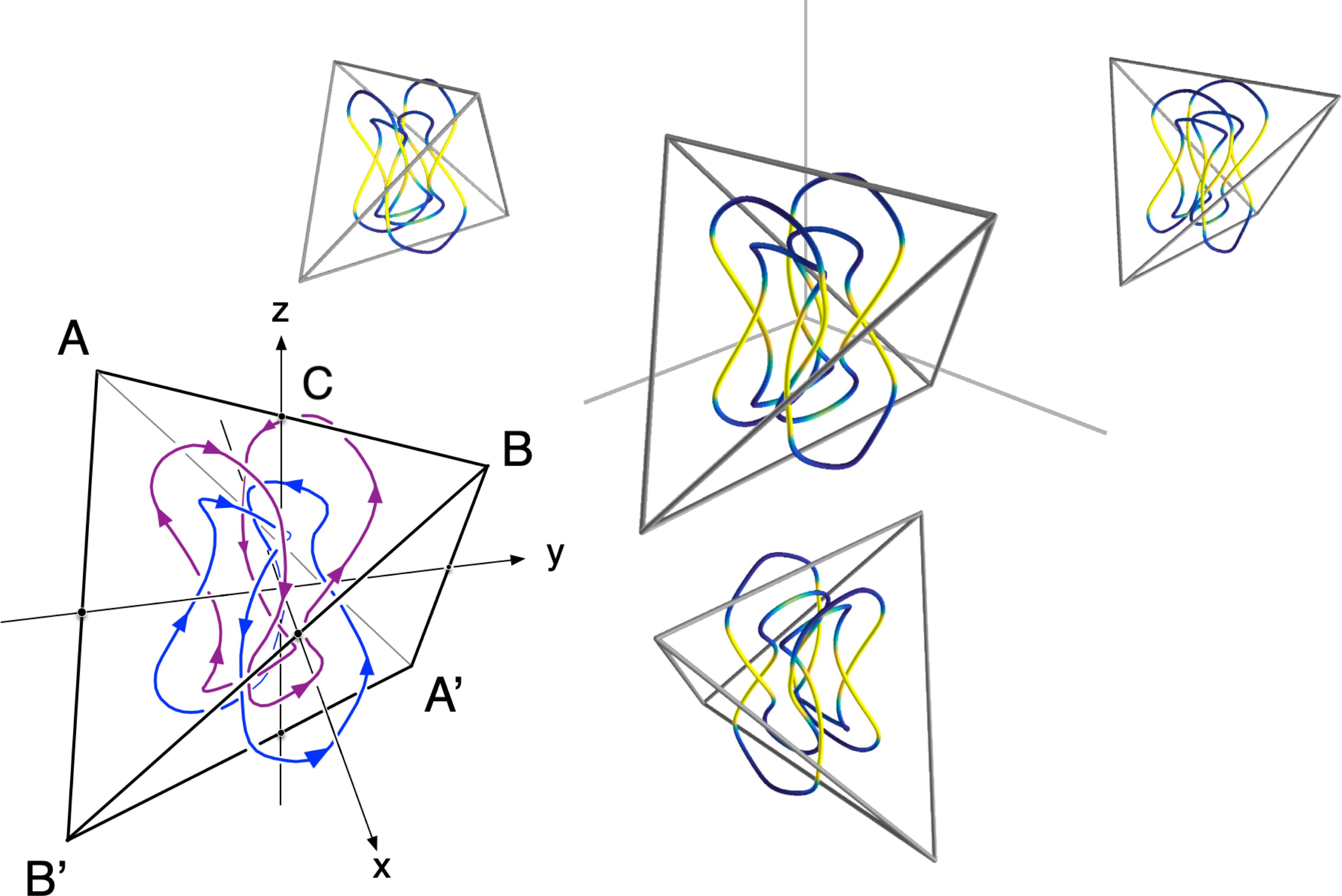}
      \caption{Magnetic field lines in the saturated state.
The field lines are traced from four seed points chosen at local maxima of the magnetic energy density in the \(z>0\) parts of the four convection cells.
The four traces form two distinct loop-like structures.}
      \label{260325094349}
\end{figure} 
The generated magnetic field also has a highly symmetric structure.
Fig.~\ref{260325094349} shows magnetic field lines traced from four seed points chosen at local maxima of the magnetic energy density in the \(z>0\) parts of the four convection cells.
The four traced lines reduce to two distinct loop-like structures: two seed points belong to one loop, and the remaining two belong to the other.
Each loop passes through all four convection cells.

\section{Symmetries of the velocity and magnetic fields}\label{sec:symmetries}
To characterize the spatial structure of the velocity and magnetic fields, we examine their symmetries under several isometries. 
Let \(\hat p\) be the reflection about the vertical plane \(y=x\),
\(\hat q\) be the reflection about the vertical plane \(y=-x\), 
and \(\hat r\) be the \(\pi\)-rotation about the \(x\) axis.
For an isometry \(\hat a\) represented by an orthogonal matrix \(R_a\), 
the induced actions on polar and axial vector fields are
\begin{align}
	(\hat a\, \bm{v})(\bm{x}) &= R_a\,\bm{v}(R_a^{-1}\bm{x}), \\
	(\hat a\, \bm{b})(\bm{x}) &= \det(R_a)\,R_a\,\bm{b}(R_a^{-1}\bm{x}).
\end{align}
The explicit matrices for them, $R_p$, $R_q$, and $R_r$, are listed in Appendix~\ref{260716101708}.
With these definitions, the saturated fields obtained by the simulation satisfy
\begin{align}
\bm{v} &= \hat{p}\,\bm{v} = \hat{q}\,\bm{v} = \hat{r}\,\bm{v}\,  \label{260512103938}\\
\bm{b} &= \hat{p}\,\bm{b} = \hat{q}\,\bm{b} = -\hat{r}\,\bm{b}\,  \label{260512103943}
\end{align}
The magnetic field is antisymmetric under the $\pi$-rotation about the $x$ axis, 
or the $\pi$-rotation about the $y$-axis ($\hat{r}\hat{p}\hat{q}$).
Defining $\hat{\sigma} = \hat{r}\hat{p}$ for the velocity field 
and $\hat{\sigma}=(-\hat{r})\hat{p}$ for the magnetic field,
where $-\hat{r}$ denotes the signed action associated with the antisymmetry of $\bm{b}$,
we have $\hat{\sigma}^4=\hat{q}^2=1$ and $\hat{q}\hat{\sigma}\hat{q}=\hat{\sigma}^{-1}$.
Thus, the symmetry group of the velocity field and the signed symmetry group of the magnetic field are both isomorphic to the dihedral group \(D_4\) of order~8.
 
We quantify the symmetry content by projecting $\bm{v}$ and $\bm{b}$ onto their $D_4$-invariant component via group averaging over the tetrahedral volume, and have found that the normalized energy fraction in the symmetric component is \(0.999\) for both fields;
see Appendix~\ref{260716101708} for operator details.

\section{Dynamo mechanism}

The schematic magnetic field in the lower-left part of Fig.~\ref{260325094349} shows the two representative field-line loops in purple and blue.
In the upper half of the tetrahedron (\(z>0\)), a portion of the purple line extends outside the fluid region, whereas the blue line lies entirely inside it; we therefore focus on the blue line in this region.
Because this line is convex upward, it exerts a downward magnetic-tension force.
The upward flow from \(B'\) toward \(C\) acts against this tension, so that \(-\bm{v}\cdot(\bm{j}\times\bm{b})>0\) and kinetic energy is converted into magnetic energy.
The same tension-work mechanism operates symmetrically for the purple line in the lower half of the tetrahedron, driven by the downward flow from \(B\).
\begin{figure}[ht]   \centering   
  \includegraphics[%
     height=1.0\textheight,%
       width=1.0\hsize,keepaspectratio]%
         {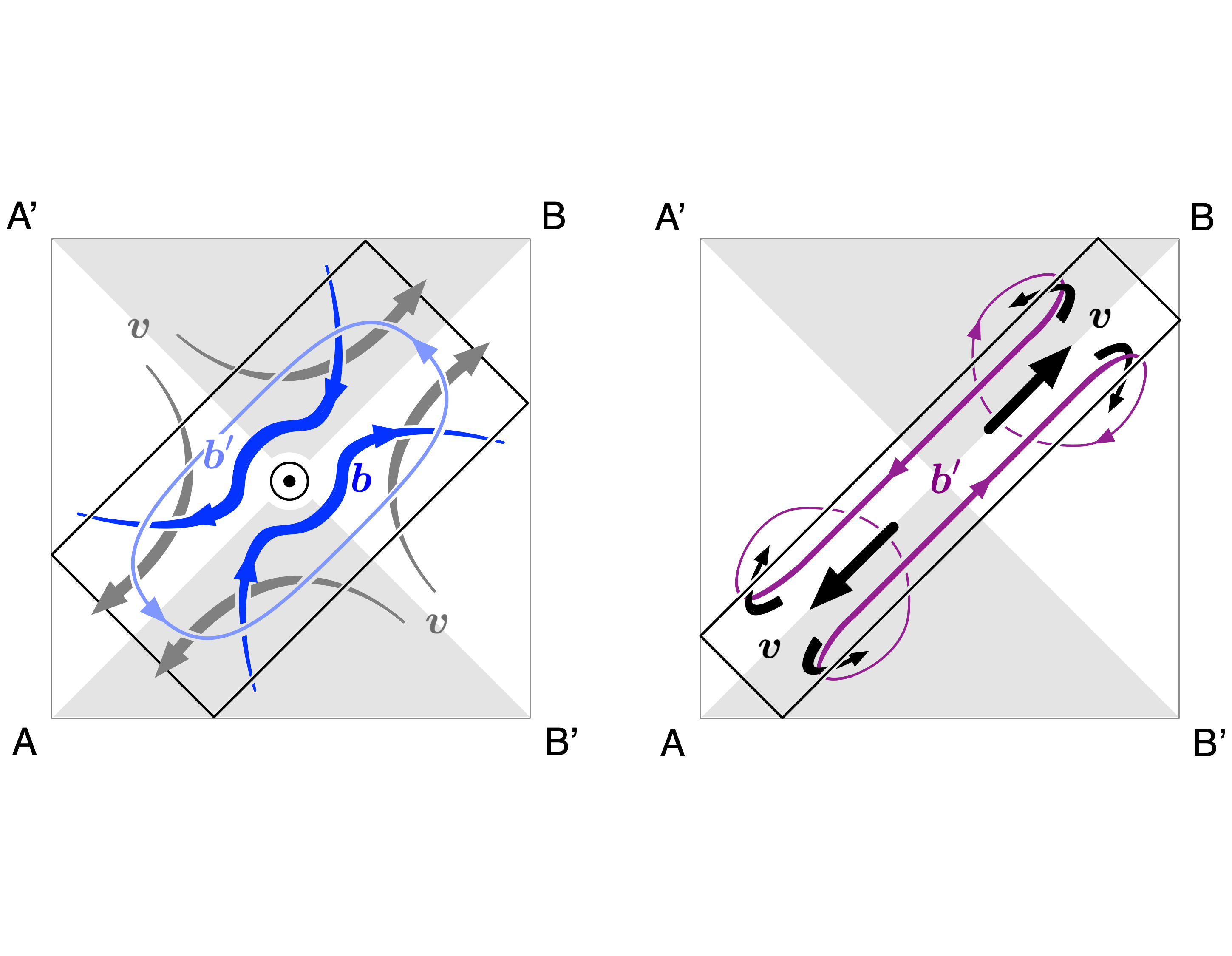}
      \caption{Schematic top views of the closure of the induction cycle at \(z\simeq 0.2\) (left) and \(z\simeq 0.3\) (right).
The square \(A'BB'A\) is the projection of the tetrahedron, and the white and gray triangles denote convection cells with positive and negative helicity, respectively.
Left: helical flows around the \(AB\)-aligned magnetic field \(\bm{b}\) induce upward currents, generating a counterclockwise horizontal field \(\bm{b}'\).
Right: the induced field is stretched and transported downward, feeding the magnetic-field component along \(A'B'\).}
      \label{260605220619}
\end{figure} 

Another energy-conversion process occurs near the upper edge \(AB\).
When the rising flow from \(B'\) reaches the vicinity of \(C\), it splits into two branches directed toward \(A\) and \(B\).
The resulting flow along \(AB\) is nearly parallel to the edge-aligned magnetic field and amplifies this component by field-line stretching.

The two mechanisms described above account for local conversion of kinetic energy into magnetic energy.
Fig.~\ref{260605220619} illustrates how these local amplification processes are connected into a global induction cycle.
The field component along \(AB\) in the \(y=x\) direction feeds the component along \(A'B'\) in the \(y=-x\) direction, and vice versa.
Helical flows in a resistive MHD fluid can induce an electric current component along the magnetic field~\cite{Moffatt2019-ew}.
The currents in the four cells, which are anti-parallel to the magnetic field in the positive-helicity cells and parallel to it in the negative-helicity cells, are all upward.
The upward currents are associated, through Amp\`ere's law, with a counterclockwise horizontal magnetic-field component \(\bm{b}'\), as shown by the pale-blue line.
This field-line loop is then stretched by oppositely directed horizontal flows along \(AB\).
The horizontal flows subsequently descend along the tetrahedral faces \(BA'B'\) and \(AA'B'\), carrying both ends of the loop into the lower half of the tetrahedron.
There, they supply the seed field for the corresponding amplification process in the \(z<0\) region.
Thus, geometrically induced helical convection closes the induction loop by mutually regenerating and stretching the two edge-aligned magnetic components along \(AB\) and \(A'B'\).

\section{Conclusion}
%
Although kinetic helicity is not a necessary condition for dynamo action, helical motion provides one important route by which conducting flows can stretch and reorient magnetic fields.
In geophysical and astrophysical settings, such motion is commonly organized by rotation.
Conventional geodynamo simulations, however, involve many coupled ingredients, including rapid rotation, spherical-shell geometry, boundary layers, turbulence, and thermal and compositional buoyancy.
The present model deliberately removes most of these ingredients and isolates a simpler question: how tetrahedral geometry organizes laminar convection and how the resulting flow closes a self-sustaining induction cycle.

In this work, we have presented a rotation-free thermal convection dynamo in a regular tetrahedron.
In the reference calculation, the saturated magnetic energy exceeds the kinetic energy of the convection.

The regular tetrahedral cavity provides a simple setting in which the organization of the laminar flow, magnetic-field amplification, and a closed induction cycle can be identified explicitly.
The high symmetry of the velocity and magnetic fields enables us to dissect the dynamo mechanism using simple field-line images.

The obtained magnetic field is antisymmetric under \(\pi\)-rotations about the two horizontal axes of the regular tetrahedron.
Together with the pair of magnetic states related by the global sign reversal \(\bm{b}\to-\bm{b}\), this geometrically constrained dynamo provides a simple setting for investigating transitions between symmetry-related magnetic states, including possible polarity reversals.

\appendix

\section{Grid-conforming placement of a regular tetrahedron}\label{260402113313}

A distinctive numerical feature of the present model is that the regular tetrahedral fluid domain can be represented \emph{exactly} on a uniform Cartesian finite-difference grid.
This avoids the need for finite-element discretizations or immersed/embedded-boundary techniques in handling oblique walls~\cite{Ghachem2021-zn,Abu-Hamdeh2022-bj,Aslam2025-zz,Dittko2013-ak}.

Let $(x,y,z)$ denote Cartesian coordinates with grid spacing $\Delta$.
We place the tetrahedron with vertices
\begin{equation} 
	A=(-c,-c,c),	\
	A'=(-c,c,-c),	\
	B=(c,c,c),		\
	B'=(c,-c,-c),
\end{equation}
where $c$ is chosen as an integer multiple of $\Delta$, i.e.\ $c=n\Delta$ with $n$ an integer.
With this choice, all vertices lie on grid points.
The four triangular faces are the planes
\begin{equation}  \label{260601180005}
	x+y-z=c,\quad
	x-y+z=c,\quad
	-x+y+z=c,\quad
	-x-y-z=c,
\end{equation}
and the tetrahedral interior satisfies the inequalities
\begin{equation}
	x+y-z < c,\quad
	x-y+z < c,\quad
	-x+y+z < c,\quad
	-x-y-z < c.
\end{equation}
On the uniform grid $x=i\Delta$, $y=j\Delta$, $z=k\Delta$ ($i,j,k\in\mathbb{Z}$), these conditions become purely \emph{integer} inequalities:
\begin{equation}  \label{260601175855}
	i+j-k < n,\quad
	i-j+k < n,\quad
	-i+j+k < n,\quad
	-i-j-k < n.
\end{equation}
Hence, each face corresponds to a set of grid points with a fixed integer value, and each edge corresponds to the intersection of two such integer planes.
In this sense, the tetrahedron is \emph{grid conforming}: 
vertices, edges, and faces are represented without geometric approximation on the underlying Cartesian mesh.

In practice,  we classify grid points into fluid-interior, wall, and exterior sets. 
Interior points satisfy all four strict inequalities in Eq.~\eqref{260601175855}, whereas wall points satisfy the corresponding non-strict inequalities with at least one equality.
All variables are stored on the full Cartesian array, but the fluid equations are advanced only at fluid-interior points, while the exterior velocity is constrained to zero.
No-slip and fixed-temperature boundary conditions are imposed at wall points satisfying any of the face equalities in Eq.~\eqref{260601180005}.
The induction equation is advanced throughout the full domain.

\section{Discrete symmetries of the velocity and magnetic fields}\label{260716101708}

The geometry-selected laminar state found in the main text is well described by a discrete symmetry group
isomorphic to the dihedral group of order~8, $D_4$, as identified in the main text.
In practice, it is convenient to specify the action of a few generating transformations on
(i)~the coordinates and (ii)~the components of the vector fields.
Below, we summarize the explicit component-wise forms used in the analysis, 
and the associated group-averaging projection onto the invariant subspace.

We use three generators, $\hat{p}$ (reflection about $y=x$), $\hat{q}$ (reflection about $y=-x$), and $\hat{r}$ ($\pi$-rotation about the $x$ axis), with
\begin{align}
	R_{p}
	=
	\begin{pmatrix}
		0&1&0\\
		1&0&0\\
		0&0&1
	\end{pmatrix},
	\
	R_{q}
	=
	    \begin{pmatrix}
		    0&-1&0\\
		    -1&0&0\\
		    0&0&1
	    \end{pmatrix},
	\
	R_{r}
	=
		\begin{pmatrix}
			1&0&0\\
			0&-1&0\\
			0&0&-1
		\end{pmatrix}.	
\end{align}
%
Let \(\hat{e}\) denote the identity transformation. 
The velocity field in the saturated state is invariant under the eight transformations,
\begin{align}		\label{260519104426}
	G_v
	=
	\left\{
		\hat{e},\ \hat{p},\ \hat{q},\ \hat{p}\hat{q},\
		\hat{r},\ \hat{r}\hat{p},\ \hat{r}\hat{q},\ \hat{r}\hat{p}\hat{q}
	\right\}.
\end{align}
The set $G_v$ forms a group isomorphic to $D_4$.

The magnetic field is invariant under the axial-vector actions associated with \(R_p\) and \(R_q\), 
whereas it changes sign under the axial-vector action associated with \(R_r\). 
Since \(\det(R_r)=1\), this antisymmetry is written as
\begin{align}
    \bm{b}(\bm{x})
    =
    -R_r\,\bm{b}(R_r^{-1}\bm{x}).
\end{align}
We define \(\hat r'\) as the signed action on the magnetic field obtained by combining the spatial isometry \(\hat r\) with the global sign reversal \(\bm b\to-\bm b\). 
The corresponding magnetic-field symmetry group is
\begin{align}
    G_b
    =
    \left\{
        \hat{e},\ \hat{p},\ \hat{q},\ \hat{p}\hat{q},\
        \hat{r}',\ \hat{r}'\hat{p},\ \hat{r}'\hat{q},\ \hat{r}'\hat{p}\hat{q}
    \right\}.
\end{align}
The set \(G_b\) forms a group isomorphic to \(D_4\).

To quantify the symmetry content, we define the \(G_v\)-symmetric component of the velocity field by
\begin{align}
    \bm{v}_{\mathrm{s}}
    =
    \hat{S}_v\bm{v},
    \quad
    \hat{S}_v
    =
    \frac{1}{8}\left(
        \hat{e} + \hat{p} + \hat{q} + \hat{p}\hat{q}
        + \hat{r} + \hat{r}\hat{p} + \hat{r}\hat{q} + \hat{r}\hat{p}\hat{q}
    \right).
\end{align}
The symmetry fraction reported in the main text is computed as
\begin{equation}
	\mathcal{F}_v
	=
	\frac{\int_{\mathcal{T}} |\bm{v}_{\mathrm{s}}|^2\, dV}{\int_{\mathcal{T}} |\bm{v}|^2\, dV},
\end{equation}
where $\mathcal{T}$ is the tetrahedral volume.

The magnetic-field symmetric component is defined analogously as
\(\bm{b}_{\mathrm{s}}=\hat{S}_b\bm{b}\), where \(\hat{S}_b=|G_b|^{-1}\sum_{g\in G_b}g\).

\begin{acknowledgments}
The author thanks Ryosuke Nakashima for valuable discussions on thermal convection and Yushiro Urano for assistance with the virtual tri-mirror visualization method.
This work was supported by JSPS KAKENHI Grant Nos. JP22K18703 and JP23K24859.
\end{acknowledgments}

%

\end{document}